\documentclass{moriond}

\usepackage{multirow}
\usepackage{footnote}

\def\Journal#1#2#3#4{{#1} {\bf #2}, #3 (#4)}

\def\PRL{\em Phys. Rev. Lett.}

\def\be{\begin{equation}}
\def\ee{\end{equation}}
\def\bea{\begin{eqnarray}}
\def\eea{\end{eqnarray}}

\newcommand{\Photo}{}

\begin{document}
\vspace*{4cm}
\title{ON THE DARK CONTENTS OF THE UNIVERSE: A EUCLID SURVEY APPROACH}

\author{ ISAAC TUTUSAUS, BRAHIM LAMINE, ALAIN BLANCHARD }

\address{Universit\'{e} de Toulouse, UPS-OMP, CNRS, IRAP, F-31028 Toulouse, France}

\maketitle\abstracts{
In this work we study the consequences of allowing non pressureless
dark matter when determining dark energy constraints. We show that
present-day dark energy constraints from low-redshift probes are extremely
degraded when allowing this dark matter variation. However, adding the
CMB we can recover the $w_{DM}=0$ case
constraints. We also show that with the future Euclid redshift survey we
expect to largely improve these constraints; but, without the complementary
information of the CMB, there is still a strong degeneracy between dark
energy and dark matter equation of state parameters.
}

\section{Introduction}
The $\Lambda$CDM cosmological model is the
current standard model in cosmology thanks to its good
phenomenological fit to cosmological data like
SNIa\,\cite{Betoule2014}, BAO\,\cite{Anderson2014} or
CMB\,\cite{Planck2015}. However, neither a dark
matter, nor a dark energy candidate have been experimentally
detected, so there is still room to study models differing from this
ideal case. There has been a great effort in studying models
accounting for different dark energy components than a cosmological
constant\,\cite{Planck2015} and there has also appeared some studies
of models accounting for variations on the dark matter
sector\,\cite{Thomas}. We
propose here to study which are the consequences on the dark energy
constraints when considering non pressureless dark matter. For this, we
use present-day data and we forecast the precision of the Euclid
redshift survey.

\section{Dark content(s) of the Universe}
In this section we use present-day data from type Ia supernovae
(SNIa), the baryonic acoustic oscillations (BAO) and the cosmic
microwave background (CMB) to constraint cosmological models with
perturbations in both the dark matter and the dark energy sector. In
all this work a flat
Universe is assumed and we consider a Robertson-Walker metric with
Friedmann-Lema\^{i}tre dynamics. 

\subsection{Method and data samples}
In the following we use compressed versions of the different probes
likelihood, since they have been shown to be faster and easier to
evaluate and still remain accurate for the most common cases. 

In order to determine the constraints on the parameters under study we
use the common $\chi^2$ minimization procedure and, since we are
combining essentially independent probes, we compute the total
$\chi^2$ function as the sum of the $\chi^2$ for each probe.

For the SNIa data we use the compressed likelihood of the JLA sample
from Betoule {\it et al.}\,\cite{Betoule2014}. Corresponding the BAO
data, we use the measurements of the ratio of the comoving sound
horizon at the redshift of baryon drag to the BAO distance scale at three different redshifts: $z=0.106$\,\cite{Beutler}, 0.35\,\cite{Pad},
0.57\,\cite{Anderson}, following Planck Collaboration XVI\,\cite{PlanckXVI}. For the CMB probe we use the values of the so-called
reduced parameters (the scaled distance to recombination $R$, the
angular scale of the sound horizon at recombination $\ell_a$ and
the reduced density parameter of the baryons $\omega_b$) provided by
the Planck 2015 data release\,\cite{Planck2015}, where
temperature-temperature fluctuations and the low-$\ell$ Planck
temperature-polarization likelihoods have been used.

\subsection{Models}
In the standard $\Lambda$CDM model there exist a dark matter component
with equation of state $P=w_{DM}\rho$, $w_{DM}=0$, and a dark energy
component with equation of state $P=w_{DE}\rho$, $w_{DE}=-1$. In this work we consider that the dark
equation of state parameters $w_{DM}$ and $w_{DE}$ are constant but
may differ from their standard value. This leads to
the Friedmann-Lema\^{i}tre equation:

\begin{equation}\label{fried}
\frac{H^2(z)}{H_0^2}=\Omega_r(1+z)^4+\Omega_b(1+z)^3+(\Omega_m-\Omega_b)(1+z)^{3(1+w_{DM})}+(1-\Omega_r-\Omega_m)(1+z)^{3(1+w_{DE})}\,.
\end{equation}

We consider three different models: the $w$CDM model, consisting of
standard cold dark matter ($w_{DM}=0$) and dark energy with constant
equation of state parameter($w_{DE}=w$); the $\epsilon$CDM model, with
constant dark matter equation of state parameter ($w_{DM}=\epsilon$)
and a cosmological constant ($w_{DE}=-1$); and the $\epsilon w$CDM
model, with both dark matter and dark energy constant equation of
state parameters ($w_{DM}=\epsilon$ and $w_{DE}=w$). Since in
the $\epsilon$CDM and $\epsilon w$CDM models we are modifying the matter component at the CMB era, we
must adapt the computation of the reduced parameters by changing the
dark matter density parameter to an effective dark matter density
parameter: $\Omega_{DM}^{eff}=\Omega_{DM}(1+z_{CMB})^{3\epsilon}$.

\subsection{Results}
The obtained constraints for the cosmological parameters\,\footnote{The
baryon density parameter is not shown because either it is fixed to
the Planck 2015\,\cite{Planck2015} value (column 1) or it is very well
constrained close to it (column 3). The radiation contribution is
fixed following Komatsu {\it et al.}\,\cite{Komatsu}.} of the
different models are shown in columns 1 and 3 of Table
\ref{table1}.

For the $w$CDM model our results
are very similar to those of Betoule {\it et
  al.}\,\cite{Betoule2014}. Concerning the
$\epsilon$CDM model we have obtained compatible results with Thomas {\it et al.}\,\cite{Thomas}. The values obtained for the $\epsilon w$CDM model are slightly worse than the ones obtained for the
$w$CDM and the $\epsilon$CDM models due to the introduction of a new
degree of freedom.

All the constraints for the proposed models are compatible with the
$\Lambda$CDM model. However, two points worth to be mentioned: first
of all, the CMB probe plays a crucial role here, since SNIa+BAO data
alone cannot constraint $\epsilon$ and $w$ at the same time. And
secondly, the constraints on dark matter and dark energy are not
completely independent (see the left panel of Fig. \ref{fig}); therefore,
all the assumptions done in one of the sectors may influence the
constraints obtained in the other one.

\section{Dark content(s) of the Universe: a Euclid forecast}
In this section we study the models departing from the $\Lambda$CDM
ideal case presented by looking at the expected precision from a
galaxy power spectrum Euclid\,\footnote{http://www.euclid-ec.org} survey
forecast. In this work we restrict
ourselves to the spectroscopic Euclid redshift survey. Adding the
photometry and the weak lensing probe we can expect even better constraints than the ones presented in this work.

\subsection{Method}
In order to perform the forecast, we use a
Fisher matrix formalism in a parameterized cosmological model,
considering the Hubble parameter and the angular-diameter distance as
observables.

The Fisher matrix in a given redshift interval is given by\,\cite{Tegmark}:

\begin{equation}\label{teg}
F_{ij}=\int_{-1}^1\int_{k_{min}}^{k_{max}}\frac{\partial \ln
  P_{obs}(k,\mu)}{\partial p_i}\frac{\partial \ln
  P_{obs}(k,\mu)}{\partial p_j}V_{eff}(k,\mu)\frac{2\pi k^2 dk d\mu}{2(2\pi)^3}\,,
\end{equation}

\noindent where $V_{eff}$ is the effective volume of the survey (redshift range:
0.7-2.1, area: 15000
sq. deg., number of galaxies: 50$\times 10^6$), $p_i$ and $p_j$
stand for the observables and $P_{obs}$ is the observed power
spectrum, which differ from the
matter power spectrum because of the biasing of galaxies and their
velocity field\,\cite{Seo}. We consider the bias given in Amendola
{\it et al.}\,\cite{Amendola}: $b(z)=\sqrt{1+z}$. We assume that there is no extra noise in
this power spectrum relation. We follow Seo \& Eisenstein\,\cite{Seo} in
cutting the integral of eq. (\ref{teg}) at $k_{min}$ and $k_{max}$ to
keep the linear part of the power spectrum. Also, we multiply the integrand
of the Fisher matrix by an exponential
suppression exp$(-k^2\mu^2\sigma_r^2)$, with
$\sigma_r=c\sigma_z/H(z)$, in order to
take into account the redshift error $\sigma_z=0.001(1+z)$ of the galaxy
survey. Once we obtain the Fisher matrix for the observables we
propagate it to the parameters under study following Wang {\it et al.}\,\cite{Wang}. The final Fisher matrix is
obtained as the sum of the different matrices for the different
redshift bins.

\subsection{Results}
The obtained constraints from the forecast on the cosmological parameters\,\footnote{The
baryon density parameter is not shown because either it is fixed to
the Planck 2015\,\cite{Planck2015} value (column 2) or it is very well constrained close to it (column 4). The radiation contribution is
fixed following Komatsu {\it et al.}\,\cite{Komatsu}.} of the
different models are shown in column 2 of Table \ref{table1}. The
constraints obtained combining the forecast with the CMB probe are
shown in column 4 of Table \ref{table1}. We have used the values
obtained in column 3 (SNIa+BAO+CMB) as fiducial model for the
forecast. 

\begin{figure}
\begin{center}
\includegraphics[scale=0.37]{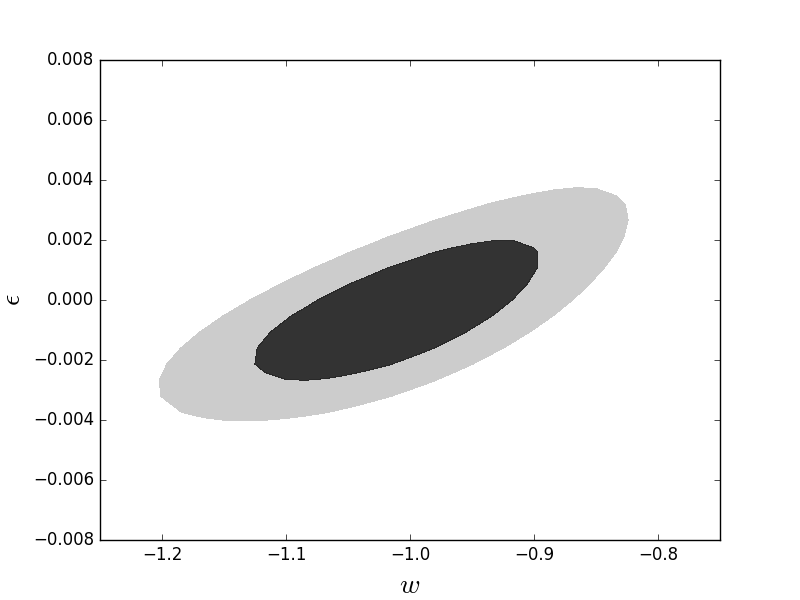}\hspace{10pt}\includegraphics[scale=0.37]{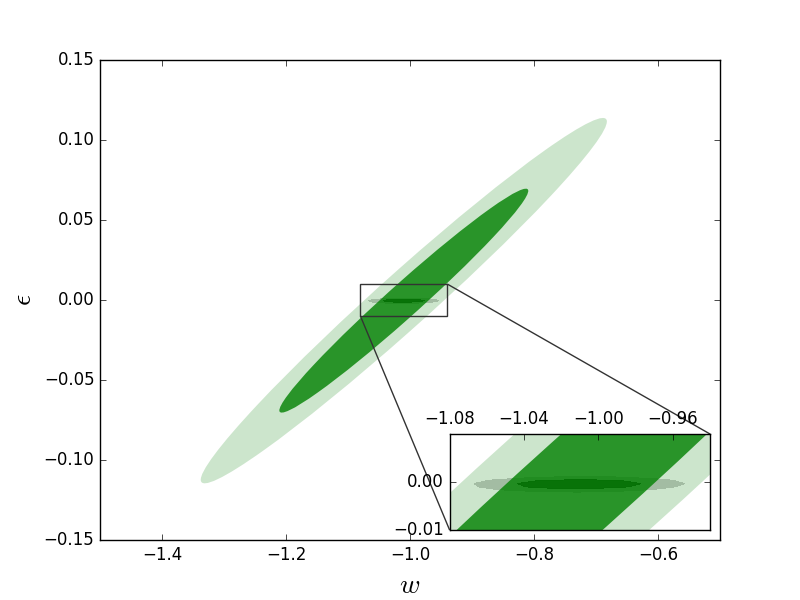}\,
\caption{Confidence contours at 68\% and 95\% confidence level for the
$w$ and $\epsilon$ parameters of the $\epsilon w$CDM model. Left
panel: SNIa+BAO+CMB present-day data. Right panel: Euclid galaxy
clustering forecast (green) and Euclid galaxy clustering+CMB (black).}\label{fig}
\end{center}
\end{figure}

For the $w$CDM and the $\epsilon$CDM models, the obtained constraints from the forecast alone are extremely better
than present-day low-redshift constraints (SNIa+BAO). When adding the
CMB to the forecast we obtain constraints between a factor 2 and 6
more precise than the current ones. Given that these results are only
for the galaxy clustering probe restricted to the linear scale we
expect even better constraints from the full exploitation of the
future Euclid data. Concerning the $\epsilon w$CDM model, the
results obtained with the forecast alone are extremely better than
SNIa+BAO data (which do not lead to significant constraints), but they
still
show a degradation on the dark energy constraints due to the dark
matter freedom. When
adding the CMB we recover constraints similar to the ones obtained for
the $w$CDM and the $\epsilon$CDM models (see right panel of Fig. \ref{fig}).

\begin{table*}[t]
\caption{Cosmological parameter constraints for the different models
  and the different probes considered. The errors are given at the
  1-$\sigma$ confidence level on 1 parameter ($\Delta \chi^2=1$). The dash in the $\epsilon w$CDM model using
SNIa+BAO data stands for the extreme degeneracies which do not allow
to obtain significant constraints on the cosmological parameters.}\label{table1}
\begin{center}
\scriptsize
\begin{tabular}{cc|c|c|c|c|}
\cline{3-6}
& & SNIa+BAO & Euclid GC & SNIa+BAO+CMB & Euclid GC + CMB\\
\cline{1-6}

\multicolumn{1}{|c}{\multirow{3}{*}{$w$CDM}} &
                                               \multicolumn{1}{|c|}{$\Omega_m$}
  & $\leq 0.28$ & $0.299\pm 0.022$ & $0.299\pm 0.012$ &
                                                             $0.2990\pm
                                                        0.0021$\\

\multicolumn{1}{|c}{} & \multicolumn{1}{|c|}{$w$} & $-0.72\pm 0.25$ &
                                                                      $-0.995\pm 0.026$ & 
                                              $-0.995\pm 0.054$
                                                           &
                                                             $-0.994\pm 0.022$\\

\multicolumn{1}{|c}{} & \multicolumn{1}{|c|}{$H_0$} & $53.0\pm 13.3$ &
                                                                       $
                                                                       68.70\pm
                                                                       0.45$
                         & $68.7\pm 1.3$ & $68.68\pm 0.40$\\

\hline

\multicolumn{1}{|c}{\multirow{3}{*}{$\epsilon$CDM}} &
                                                      \multicolumn{1}{|c|}{$\Omega_m$}
  & $\geq 0.31$ & $0.301\pm 0.010$ & $0.301\pm 0.014$ &
                                                             $0.3001\pm
                                                        0.0030$\\

\multicolumn{1}{|c}{} & \multicolumn{1}{|c|}{$\epsilon$} & $-0.49\pm
                                                           0.44$ &
                                                                   $-0.0003\pm 0.0092$ & 
                                              $-0.0003\pm 0.0011$
                                                           &
                                                             $-0.00024\pm 0.00066$\\

\multicolumn{1}{|c}{} & \multicolumn{1}{|c|}{$H_0$} & $50.00\pm 3.83$
             & $ 68.60\pm 0.27$ & $68.6\pm 1.2$ & $68.62\pm 0.12$\\

\hline

\multicolumn{1}{|c}{\multirow{4}{*}{$\epsilon w$CDM}} &
                                                        \multicolumn{1}{|c|}{$\Omega_m$}
  & & $0.301\pm 0.041$ & $0.301\pm 0.014$ &
                                                             $0.3011\pm
                                            0.0038$\\

\multicolumn{1}{|c}{} & \multicolumn{1}{|c|}{$w$} &  & $-1.01\pm 0.13$ & 
                                              $-1.010\pm 0.077$
                                                           &
                                                             $-1.010\pm 0.023$\\

\multicolumn{1}{|c}{} & \multicolumn{1}{|c|}{$\epsilon$} & $-$ &
                                                                 $0.000\pm 0.046$ & 
                                              $-0.0004\pm 0.0016$
                                                           &
                                                             $-0.00045\pm 0.00066$\\

\multicolumn{1}{|c}{} & \multicolumn{1}{|c|}{$H_0$} &  & $ 68.6\pm
                                                         1.0$ &
                                                                $68.6\pm
                                                                1.3$ &
                                                                       $68.60\pm 0.44$\\

\hline

\end{tabular}
\end{center}
\end{table*}

\section{Conclusions}
We have examined the consequences on the obtained constraints of dark
energy properties when the pressureless dark matter assumption is
relaxed (assuming constant
equation of state parameter for both dark matter and the dark
energy). We have found that for low redshift present-day data this freedom in
the dark matter equation of state parameter completely degrades the
constraints on the dark energy sector. When adding the CMB we almost
recover the pressureless dark matter case constraints thanks to the tight
constraint on the dark matter equation of state parameter from the
CMB. We have also seen that the galaxy clustering probe of the Euclid
survey will perform extremely better than low redshift present-day
data, but in order to reach the same precision as in the pressureless
dark matter case we need to add the CMB. This strong role from the CMB
probably comes from assuming a constant equation of state parameter up
to the CMB redshift, but focusing in low redshift data or general
variations of the dark matter sector we will have to deal with the
observed dark matter and dark energy equation of state parameter degeneracy.


\section*{References}

\begin{thebibliography}{99}
\bibitem{Betoule2014} M. Betoule {\it et al.},
  \Journal{Astron. Astrophys.}{568}{A22}{2014}.

\bibitem{Anderson2014} L. Anderson {\it et al.}, \Journal{Mon. Not. Roy. Ast. Soc.}{441}{24}{2014}. 

\bibitem{Planck2015} Planck Collaboration 2015, arXiv:1502.01590
  [astro-ph.CO] 

\bibitem{Thomas} D. B. Thomas {\it et al.} 2016, arXiv:1601.05097
  [astro-ph.CO] 

\bibitem{Komatsu} E. Komatsu {\it et al.},
  \Journal{ApJS}{192}{18}{2011}.

\bibitem{Beutler} F. Beutler {\it et al.},
  \Journal{Mon. Not. Roy. Ast. Soc.}{416}{3017}{2011}.

\bibitem{Pad} N. Padmanabhan {\it et al.},
  \Journal{Mon. Not. Roy. Ast. Soc.}{427}{2132}{2012}.

\bibitem{Anderson} L. Anderson {\it et al.},
  \Journal{Mon. Not. Roy. Ast. Soc.}{427}{3435}{2012}.

\bibitem{PlanckXVI} Planck Collaboration, \Journal{Astron. Astrophys.}{571}{A16}{2014}.

\bibitem{Tegmark} M. Tegmark, \Journal{\PRL}{79}{3806}{1997}.

\bibitem{Seo} H. Seo \& D. J. Eisenstein, \Journal{Astrophys. J.}{598}{720}{2003}.

\bibitem{Amendola} L. Amendola {\it et al.}, \Journal{Liv. Rev. in Rel.}{16}{6}{2013}.

\bibitem{Wang} Y. Wang {\it et al.}, \Journal{Mon. Not. Roy. Ast. Soc.}{409}{737}{2010}.

\end{thebibliography}
\end{document}